\documentclass[2p]{elsarticle}
\usepackage[utf8]{inputenc}
\usepackage{natbib}
\usepackage{amsmath}
\usepackage{mathtools}
\usepackage{commath}
\usepackage{amsfonts}
\usepackage{amssymb}
\usepackage{graphicx}
\usepackage{tabularx}
\usepackage{booktabs}
\usepackage{siunitx}
\usepackage{wrapfig}
\usepackage[version=3]{mhchem}
\usepackage{subfigure}
\usepackage{float}
\usepackage{multirow}
\usepackage{caption}
\usepackage{xcolor}
\biboptions{sort&compress}	
\usepackage{fancyhdr}
\newenvironment{keywords}%
   {\begin{trivlist}\item[]{\bfseries\sffamily Keywords:}\ }
   {\end{trivlist}}

\journal{Springer, License CC-BY-NC-ND 4.0}
\title{Design of an Intake and a Thruster for an Atmosphere-Breathing Electric Propulsion System}

\author[epfl,irs]{F.~Romano\corref{cor1}\fnref{fn1}}
\ead{francesco.romano@epfl.ch}

\author[irs]{G.~Herdrich} 
\author[irs]{Y.-A.~Chan}
\author[UniMAN]{N.H.~Crisp}

\author[UniMAN]{P.C.E.~Roberts}
\author[UniMAN]{B.~E.A.~Holmes}
\author[UniMAN]{S.~Edmondson}
\author[UniMAN]{S.~Haigh}
\author[UniMAN]{A.~Macario-Rojas}
\author[UniMAN]{V.T.~A.~Oiko}
\author[UniMAN]{L.A.~Sinpetru}
\author[UniMAN]{K.~Smith}

\author[Deimos]{J.~Becedas}
\author[Deimos]{V.~Sulliotti-Linner}

\author[GomSpace]{M.~Bisgaard}
\author[GomSpace]{S.~Christensen}
\author[GomSpace]{V.~Hanessian}
\author[GomSpace]{T.~Kauffman Jensen}
\author[GomSpace]{J.~Nielsen}

\author[irs]{S.~Fasoulas}
\author[irs]{C.~Traub}

\author[UPC]{D.~García-Almiñana}
\author[UPC]{S.~Rodríguez-Donaire}
\author[UPC]{M.~Sureda}

\author[UCL]{D.~Kataria}

\author[Eurocons]{B.~Belkouchi}
\author[Eurocons]{A.~Conte}
\author[Eurocons]{S.~Seminari}
\author[Eurocons]{R.~Villain}

\fntext[fn1]{Postdoctoral Researcher, francesco.romano@epfl.ch}

\address[epfl]{Ecole Polytechnique Fédérale de Lausanne (EPFL), Swiss Plasma Center (SPC), CH-1015 Lausanne, Switzerland}
\address[irs]{Institute of Space Systems (IRS), University of Stuttgart, 70569, Germany}
\address[UniMAN]{The University of Manchester, Oxford Road, Manchester, M13 9PL, UK}
\address[Deimos]{Elecnor Deimos Satellite Systems, C/ Francia 9, 13500, Puertollano, Spain}
\address[GomSpace]{GomSpace AS, Langagervej 6, Aalborg East 9220, Denmark}
\address[UPC]{UPC-BarcelonaTECH, Colom 11, TR5 - 08222 Terrassa, Spain}
\address[UCL]{Mullard Space Science Laboratory, University College London, Holmbury St.~Mary, Dorking, Surrey, RH5 6NT, UK}
\address[Eurocons]{Euroconsult, 86 Boulevard de Sébastopol, 75003 Paris, France}
\begin{document}

\begin{abstract}
Challenging space missions include those at very low altitudes, where the atmosphere is the source of aerodynamic drag on the spacecraft, that finally defines the mission's lifetime, unless a way to compensate for it is provided. This environment is named Very Low Earth Orbit (VLEO) and it is defined for $h<\SI{450}{\kilo\meter}$. In addition to the spacecraft's aerodynamic design, to extend the lifetime of such missions, an efficient propulsion system is required.

One solution is Atmosphere-Breathing Electric Propulsion (ABEP), in which the propulsion system collects the atmospheric particles to be used as propellant for an electric thruster. The system could remove the requirement of carrying propellant on-board, and could also be applied to any planetary body with atmosphere, enabling new missions at low altitude ranges for longer missions' duration. One of the objectives of the H2020 DISCOVERER project, is the development of an intake and an electrode-less plasma thruster for an ABEP system.

This article describes the characteristics of intake design and the respective final designs based on simulations, providing collection efficiencies up to $94\%$. Furthermore, the radio frequency (RF) Helicon-based plasma thruster (IPT) is hereby presented as well, while its performances are being evaluated, the IPT has been operated with single atmospheric species as propellant, and has highlighted very low input power requirement for operation at comparable mass flow rates $P\sim\SI{60}{\watt}$.

\end{abstract}
 \maketitle

\section*{Nomenclature}
\noindent
ABEP: Atmosphere-Breathing Electric Propulsion\\
DSMC: Direct Simulation Monte Carlo \\
FMF: Free Molecular Flow \\
GSI: Gas-Surface Interaction \\
IPT: RF Helicon-based Plasma Thruster\\
VLEO: Very Low Earth Orbit\\
SC: Spacecraft 

\section{Introduction} 
\subsection{ABEP Concept}
\begin{figure}[H]
	\centering
	\includegraphics[width=12.5cm]{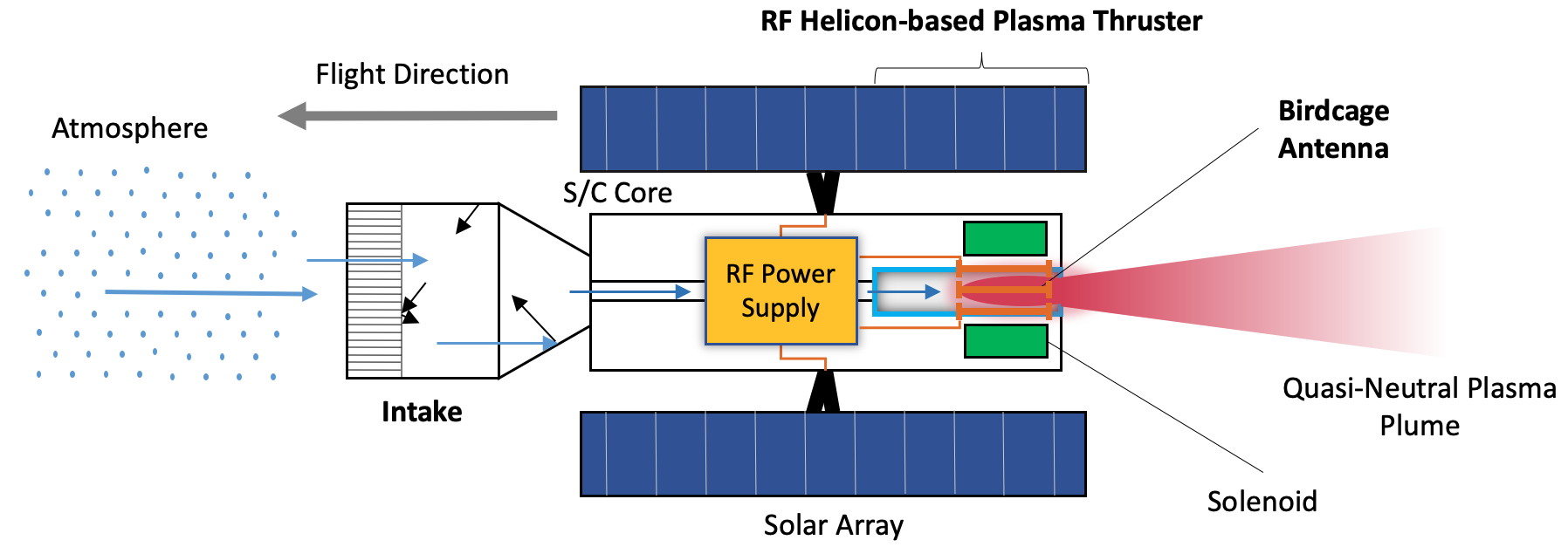}
	\caption{Atmosphere-Breathing Electric Propulsion Concept using the RF Helicon-based Plasma Thruster (IPT)~\cite{romanoacta2}.}
	\label{fig:ABEP}
\end{figure}

An atmosphere-breathing electric propulsion system (ABEP), see Fig.~\ref{fig:ABEP}, is mainly made of two components: the intake and the electric thruster. The system is designed for spacecrafts (SCs) orbiting at very low altitude orbits, for example in very low Earth orbit (VLEO), defined for altitudes $h<~\SI{450}{\kilo\meter}$~\cite{vleobenefit}. The ABEP system collects the residual atmospheric particles encountered by the SC through the intake and uses them as propellant for the electric thruster. The system is theoretically applicable to any planetary body with an atmosphere, and can drastically reduce the on-board propellant storage requirement, while extending the mission's lifetime~\cite{romanoacta}. Many ABEP concepts have been investigated in the past based on radio frequency ion thrusters (RIT)~\cite{di2007ram,presitael1,presitael2,TSAGI1,TSAGI2,filatyev2019control,TSaGI2018a,TSAGI2018ab,erofeev2017air,kanev2015electro}, ECR-based thruster~\cite{JAXA,JAXA2,JAXA3,JAXA4,JAXA5}, Hall-effect thrusters (HET)~\cite{busek,busek2,SITAEL2015,SITAEL2016,SITAEL2017,SITAEL2019a,SITAEL2019b,bauman}, and plasma thrusters~\cite{shabshelowitz2013study}. The only laboratory tested ABEP systems to date are the ABIE developed in Japan, composed of an annular intake and an ECR-based thruster into one device~\cite{JAXA,JAXA2,JAXA3,JAXA4}, and the RAM-HET system developed in Europe, comprised of the intake and a HET assembled into one device~\cite{SITAEL2015,SITAEL2016,SITAEL2017,SITAEL2019a,SITAEL2019b}. Current research in Europe is developed within the H2020 DISCOVERER and AETHER projects. 

\subsection{Very Low Earth Orbit}
\begin{figure}[H]
 \centering
 \includegraphics[width=.7\textwidth, trim={0cm 0cm 0cm 1cm},clip]{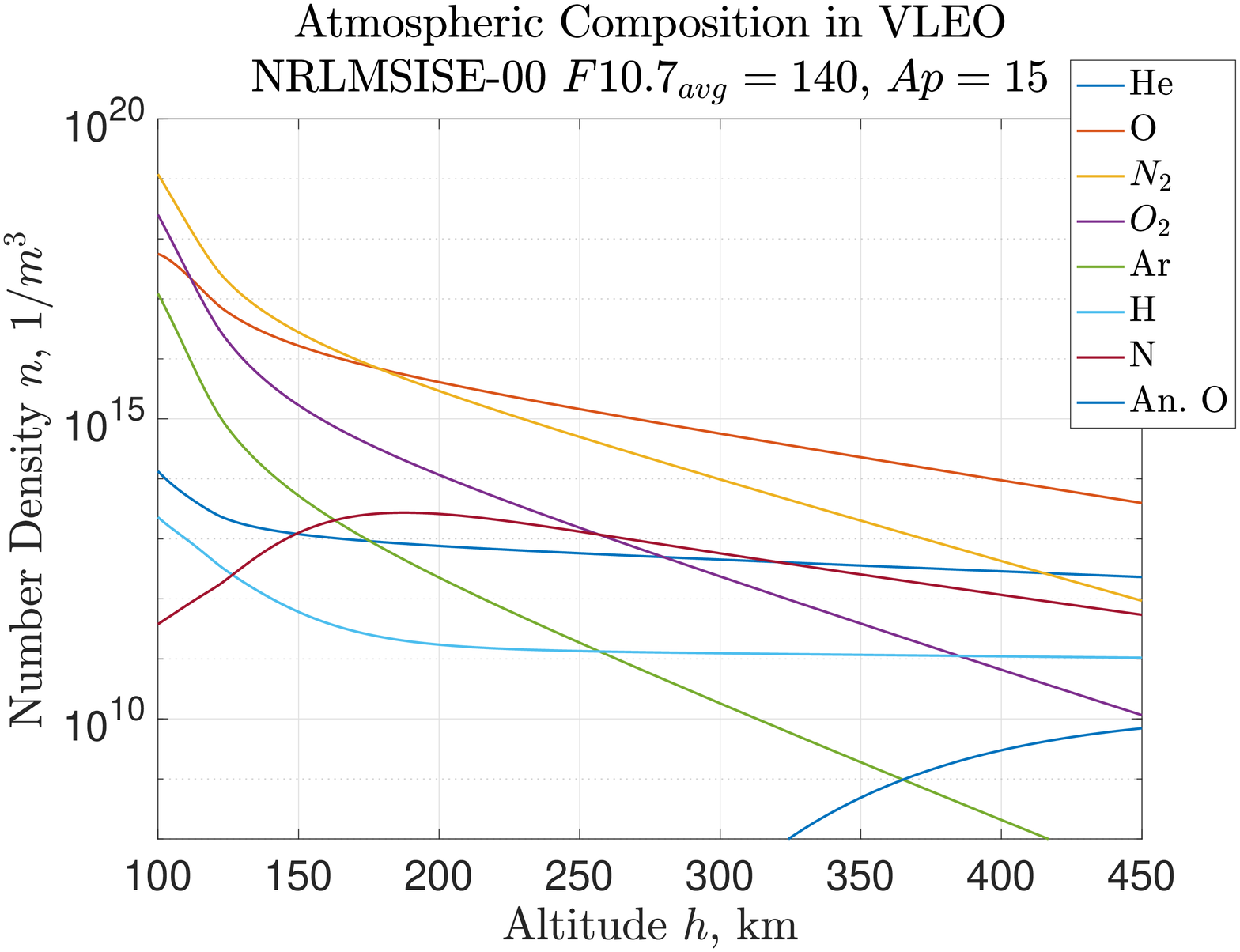}
 \caption{Atmospheric Composition in VLEO, NRLMSISE-00 Atmospheric Model: $F10.7 = 140$, $Ap = 15$, Moderate Solar Activity.}
 \label{fig:nrlmsise}
\end{figure}

The ABEP operational environment is the VLEO altitude range, that can be described by mainly density and composition over altitude and location based on atmospheric models. Among the several available, the NRLMSISE-00 is used as it accurately accounts for the lower thermosphere atmospheric conditions, as well as for being used for previous work on ABEP systems~\cite{VLEO_LEOMANNI2017,6945885,romanoacta}. 
In Fig.~\ref{fig:nrlmsise}, the averaged particle number densities are plotted in VLEO for an equatorial orbit - an aspect that fosters relevant verifications on level of the comparison between different studies. 
The solar activity is set as moderate with a respective $F10.7 = 140$, the solar radio flux at a wavelength of $\lambda=\SI{10.7}{\centi\meter}$, and the geomagnetic index $Ap=15$~\label{sec:NRLMSISE}.  The ABEP altitudes range is set to $h=150-\SI{250}{\kilo\meter}$. The lower limit is due to aerodynamic drag, as it increases exponentially leading to large power requirements~\cite{6945885} and also heating of the SC~\cite{romanoacta}. The upper limit is set by the limited propellant collection, that can become lower than conventional thruster requirements by a typical sized SC, and is where a conventional electric propulsion system becomes preferable against an ABEP system~\cite{di2007ram}. The dominant gas species in the selected altitude range are \ce{N2} and atomic oxygen (AO).

It is very important to mention, that an ABEP system must be able to perform in a wide operational envelope of parameters. Indeed, a variable incoming propellant composition and flow is expected, that also translates in larger or lower values of aerodynamic drag. This is in particular due to the non-uniformity of VLEO environment because of temporal (time), seasonal (solar activity), and geographical (latitude and longitude) variations. Also, one must take into account for more expectional space weather events, such high energetic particle fluxes coming from the Sun due to, for example, UV or X-ray flares. Such effects would change the drag, therefore the thrust requirements, as well as the propellant composition and, therefore, the power requirements, and need to be taken into account for the system's design. A more detailed analysis can be found in~\cite{romano_2022,vaidya2022development}.

\section{ABEP Intake}
\subsection{Intake Working Principle}
\begin{figure}[H]
	\centering
	\includegraphics[width=10cm]{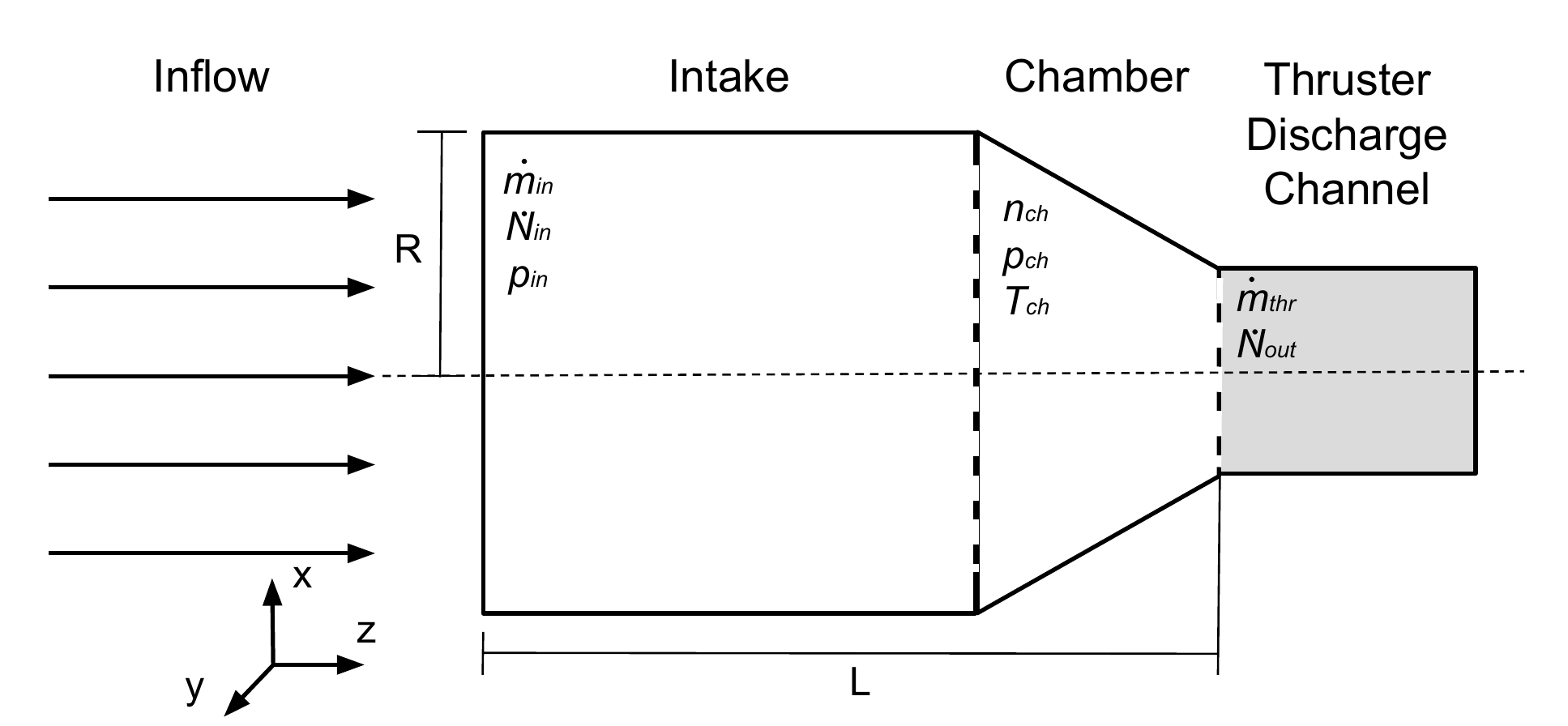}
	\caption{Intake General Design including the Thruster's Discharge Channel~\cite{romanoacta3}.}
	\label{fig:Gen_Diag}
\end{figure}

The intake is the device of an ABEP system that collects and drives the atmospheric particles to the thruster. The general schematics is shown in Fig.~\ref{fig:Gen_Diag}. The intake shall ensure the efficient collection of atmospheric particles and provide the required conditions of density, pressure, and mass flow for thruster’s operation while, at the same time, minimizing the required frontal area and, therefore, the resulting drag.

The number density is $n$ and $\dot{N}$ is the particle flux. 
The collection efficiency $\eta_c(h)$ is defined in Eq.~\ref{Eq:intake_eff}, where $\dot{N}_{out}$ is the collected particle flux, and $\dot{N}_{in}$ is the incoming one. The subscript $i$ refers to each different particle species $i= [1, N_s]$, where $N_s$ is the total number of species.
\begin{equation}
 \eta_c (h) =  \frac{\sum_{i=1}^{N_s}\dot{N}_{out_i}}{\sum_{i=1}^{N_s}\dot{N}_{in_i}}
 \label{Eq:intake_eff}
\end{equation}

The resulting mass flow to the thruster $\dot{m}_{thr}$ is obtained as shown in Eq.~\ref{Eq:mass_flow}, where $m_{p}$ is the particle mass.

\begin{equation}
 \dot{m}_{thr}(h) = \sum_{i=1}^{N_s} m_{p_i} ~ \dot{N}_{out_i} (h)
 \label{Eq:mass_flow}
\end{equation}

It shall be noted that each intake design needs to be tailored to the thruster applied, its requirements, and the mission profile. The latter, in terms of orbit (altitude and inclination), and total mission duration to account for solar activity variations.

Intake design, due to the free molecular flow condition in VLEO, is mostly driven by gas-surface interaction (GSI) properties of the materials~\cite{LIVADIOTTI2020}. Based on the Maxwell's model, upon particle impact with a surface, an exchange of momentum and energy takes place. In this model, two types of surface-particle interactions are modelled: specular and diffuse reflections. In specular reflections, particles are reflected without energy exchange, the angle of reflection equals the incident angle, and the velocity vector undergoes a complete inversion. In diffuse reflection, instead, particles reach thermal equilibrium with the surface and are reflected according to a Rayleigh distribution corresponding to the wall temperature $T_{wall}$ with only thermal velocity.

The accommodation coefficient $\alpha$ is defined in Eq.~\ref{Eq:sigmaB}, where the energy of the incident particle is $E$, and $E'$ the energy of the reflected particle, while $E_W$ is the energy which a particle would have after full accommodation to $T_{wall}$. For specular reflections, $\alpha=1$, while for diffuse ones is $\alpha=0$.
\begin{equation}
 \alpha = \frac{E-E'}{E - E_W}
 \label{Eq:sigmaB}
\end{equation}

The intakes hereby presented have been designed using the PICLas tool, which is a three-dimensional numerical tool for simulating non-equilibrium gas and plasma flows~\cite{Munz2014662,PICLAS}. For the intake design, the DSMC module with the Maxwell model for surface interactions is used~\cite{romanoacta3}. Details on the inflow parameters, that are based on NRLMSISE-00 model at moderate solar activity, can be found in~\cite{romanoacta3,romano_2022}.

\subsection{Diffuse Intake}
\begin{figure}[H]
    \centering
    \includegraphics[width=0.7\textwidth, trim={3cm 0cm 3cm 3cm},clip]{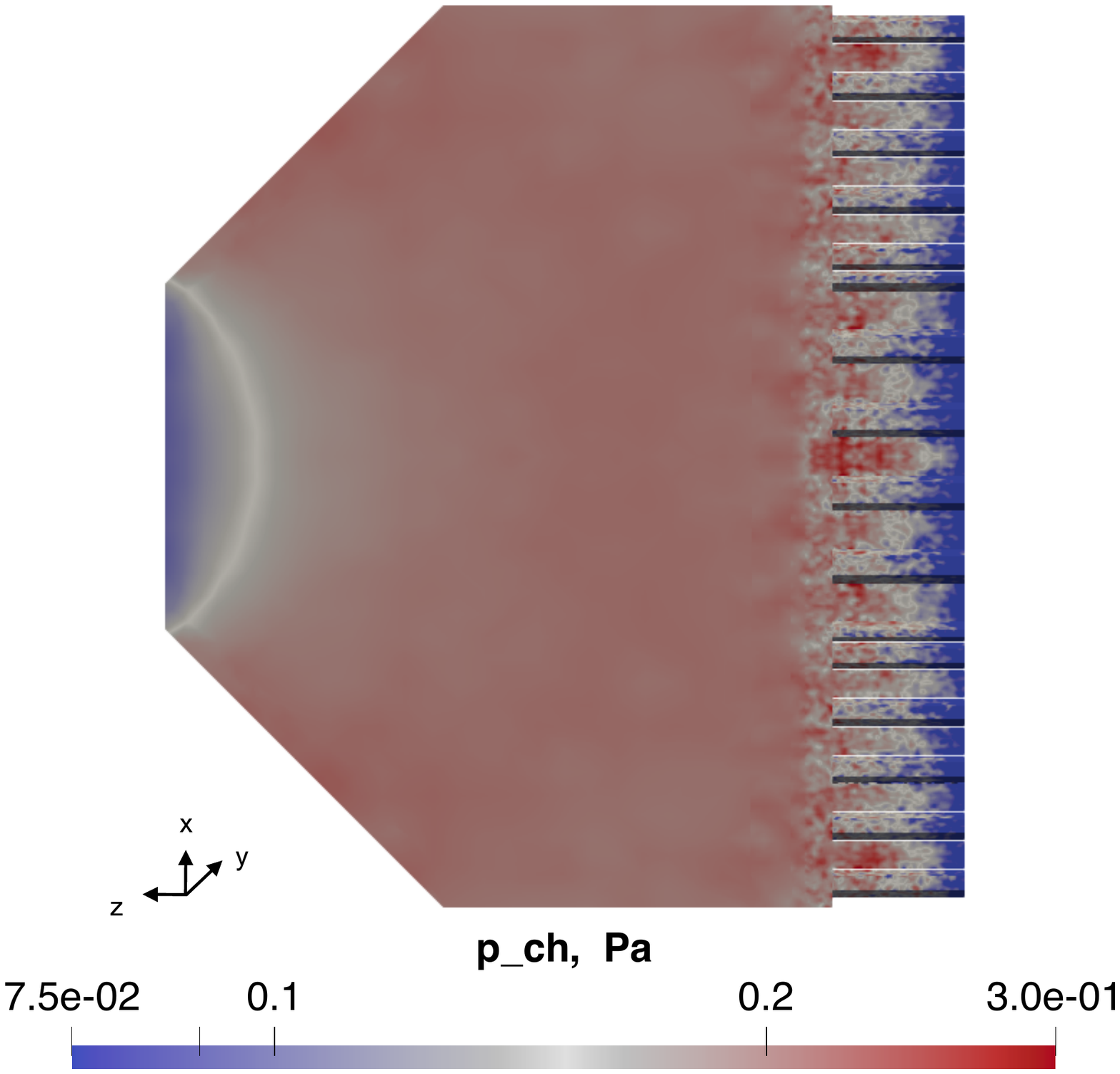}
    \caption{Diffuse Intake Pressure Distribution~\cite{romanoacta3}.}
    \label{c6_pres}
\end{figure}

\begin{figure}[H]
    \centering
    \includegraphics[width=.55\textwidth, trim={13cm 6.5cm 16cm 5cm},clip]{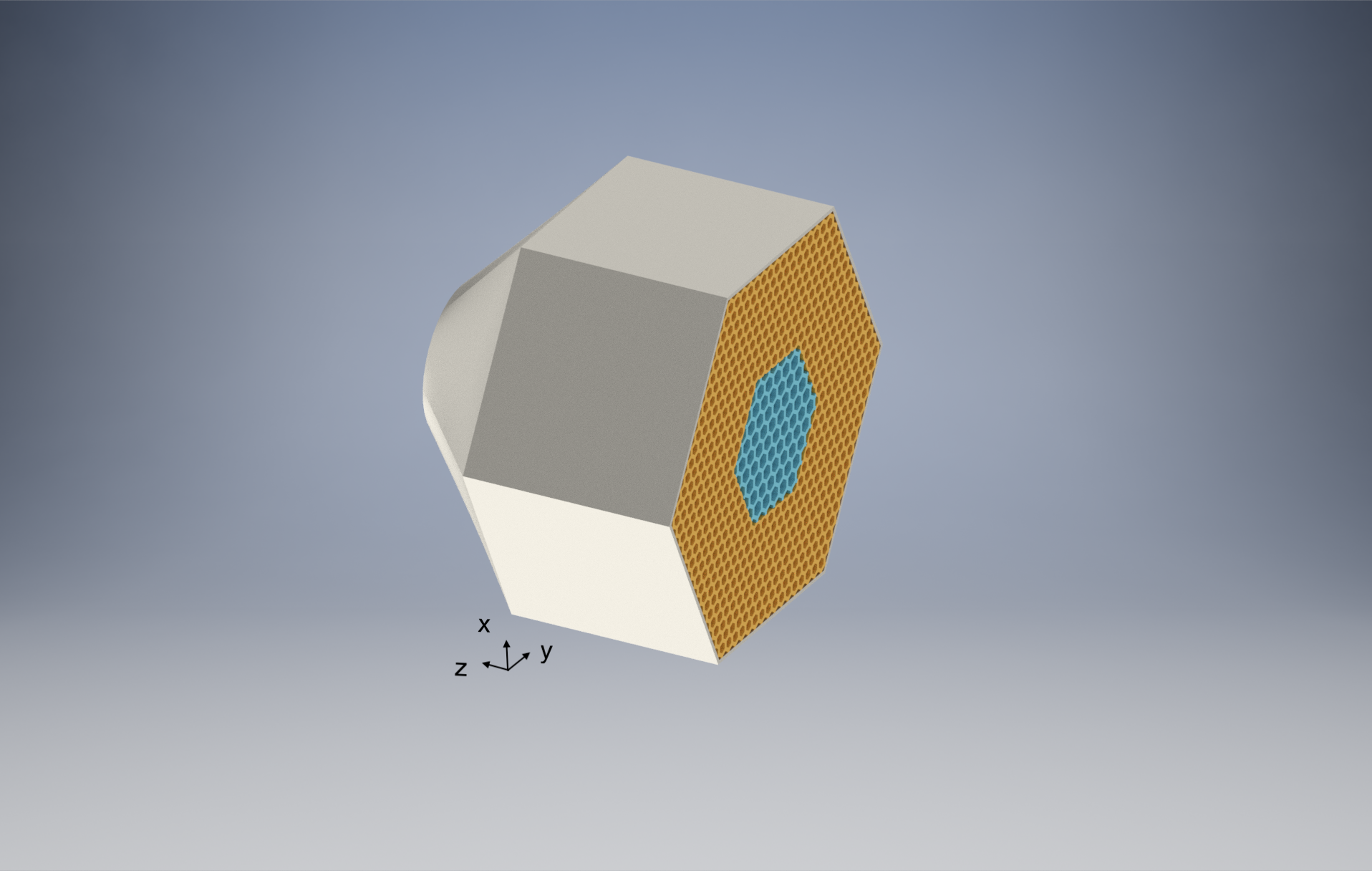}
    \caption{Diffuse Intake Render~\cite{romanoacta3}.}
    \label{dif_intake_sch}
\end{figure}

Based on the analysis of the intake concepts using diffuse scattering $\alpha=0$, this design provides a homogenous pressure distribution in front of the thruster's discharge channel as shown in Fig.~\ref{c6_pres}. The intake has an hexagonal shape and it features a front section of hexagonal ducts with different area ratio $AR$: larger ducts in front of the thruster's discharge channel to increase $A_{in}$, and narrower ones in the outer intake region to  reduce the backflow. At the rear there is a conical section with a chamber angle of \SI{45}{\degree} terminating on the thruster's discharge channel diameter. In Fig.~\ref{dif_intake_sch} the isometric render view of the diffuse intake is shown. This design provides a theoretical maximum of $\eta_c=0.458$~\cite{romanoacta3}.

\subsection{Specular Intake}
\begin{figure}[H]
    \centering
    \includegraphics[width=.7\textwidth, trim={9cm 2cm 8cm 2cm},clip]{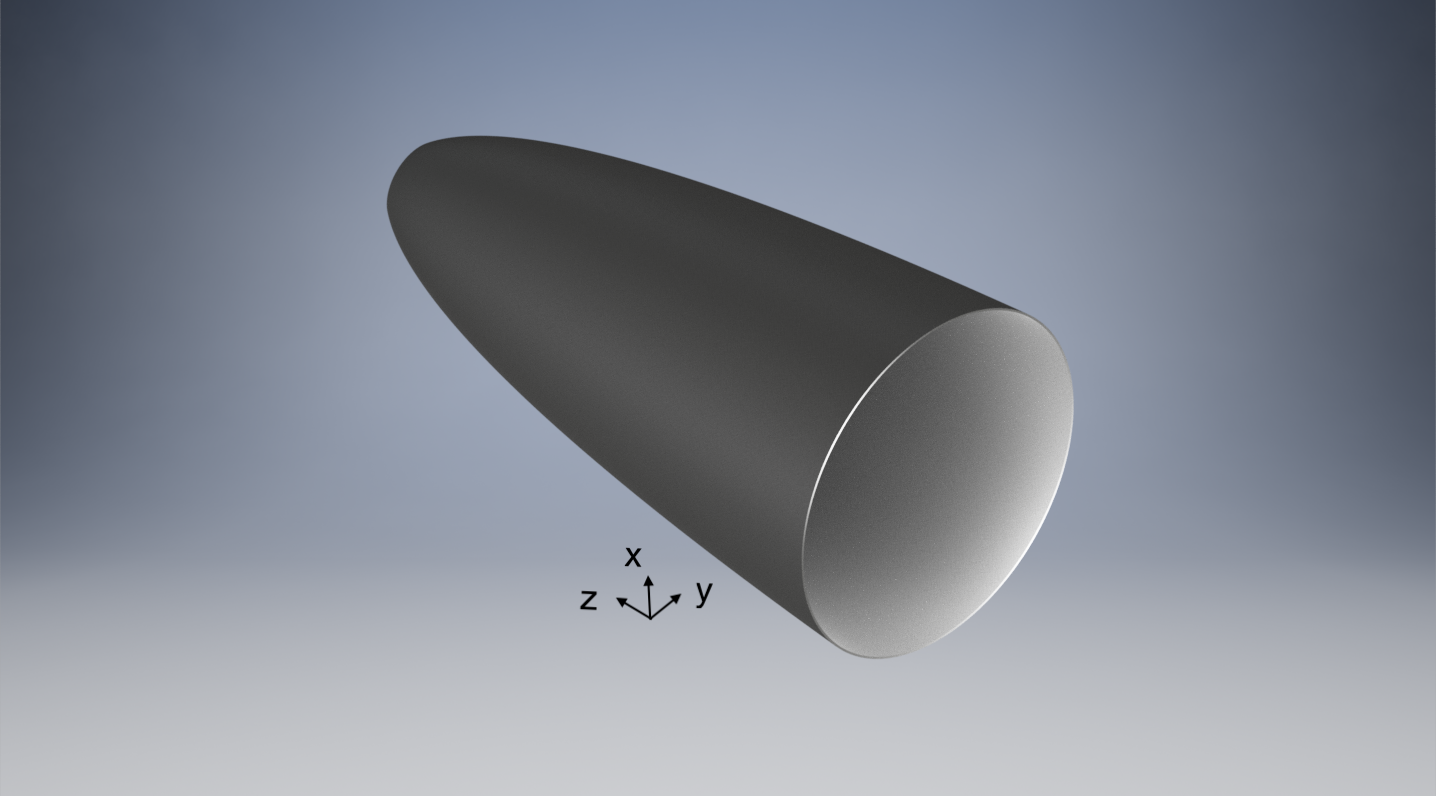}
    \caption{Specular Intake Render~\cite{romanoacta3}.}
    \label{opt_intake_rend}
\end{figure}

Based on specular scattering properties $\alpha=1$, a parabolic scoop with its focus inside the thruster discharge channel provides a theoretical $\eta_c=0.94$~\cite{romanoacta3}. This design also outperforms the diffuse intake in terms of sensitivity to flow misalignment leading to a limit of $\alpha=\SI{15}{\degree}$ resulting in $\eta_c<0.6$. The isometric render view of the specular intake is shown in Fig.~\ref{opt_intake_rend}~\cite{romanoacta3}.
\begin{figure}[H]
 \centering
 \includegraphics[width=.95\textwidth, trim={0cm 1cm 0cm 3cm},clip]{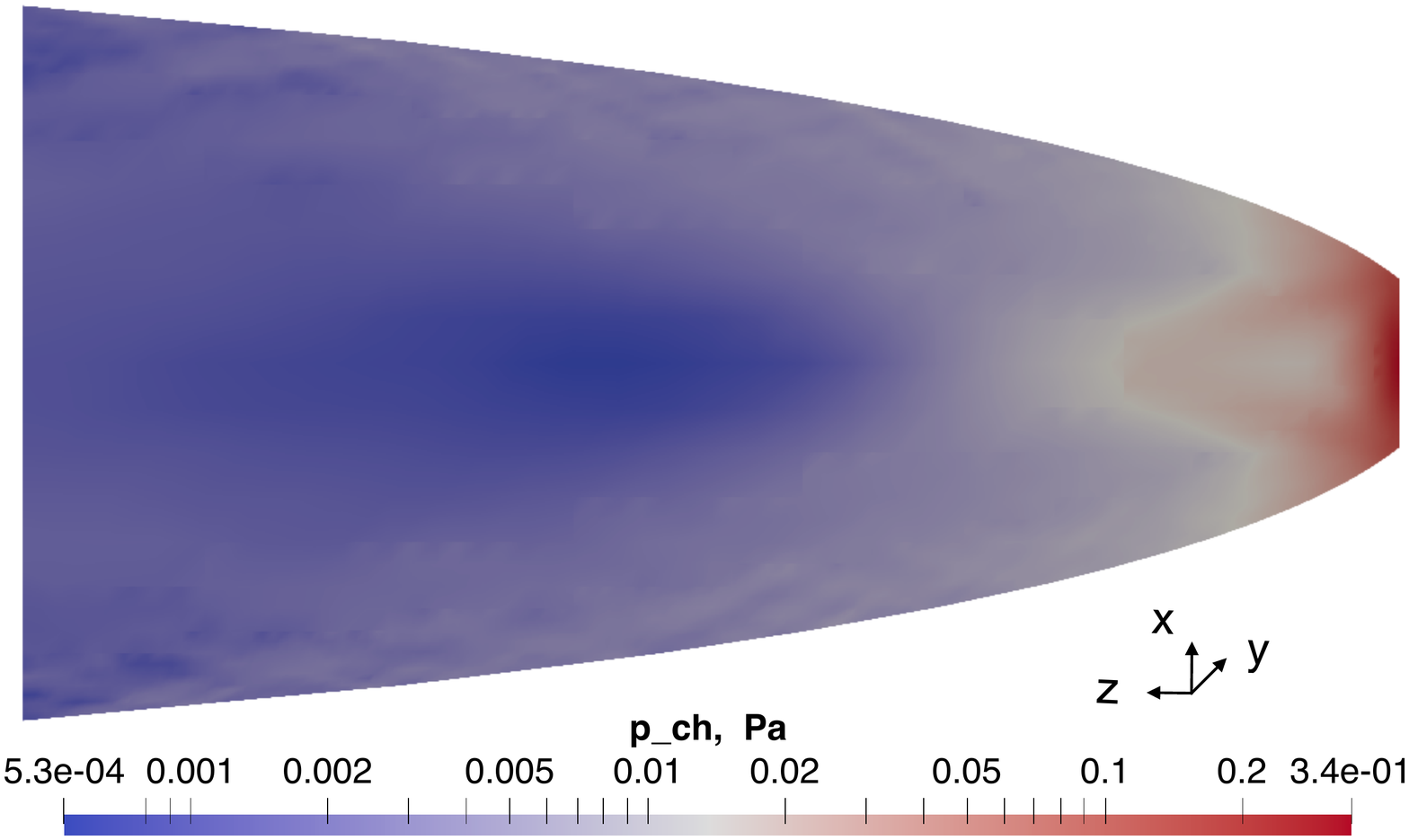}
 \caption{Specular Intake Pressure Distribution~\cite{romanoacta3}.}
 \label{c10-press}
\end{figure}

\section{ABEP Thruster}

\subsection{Thruster Working Principle}
\begin{figure}[H]
    \centering
    \includegraphics[width=.9\textwidth]{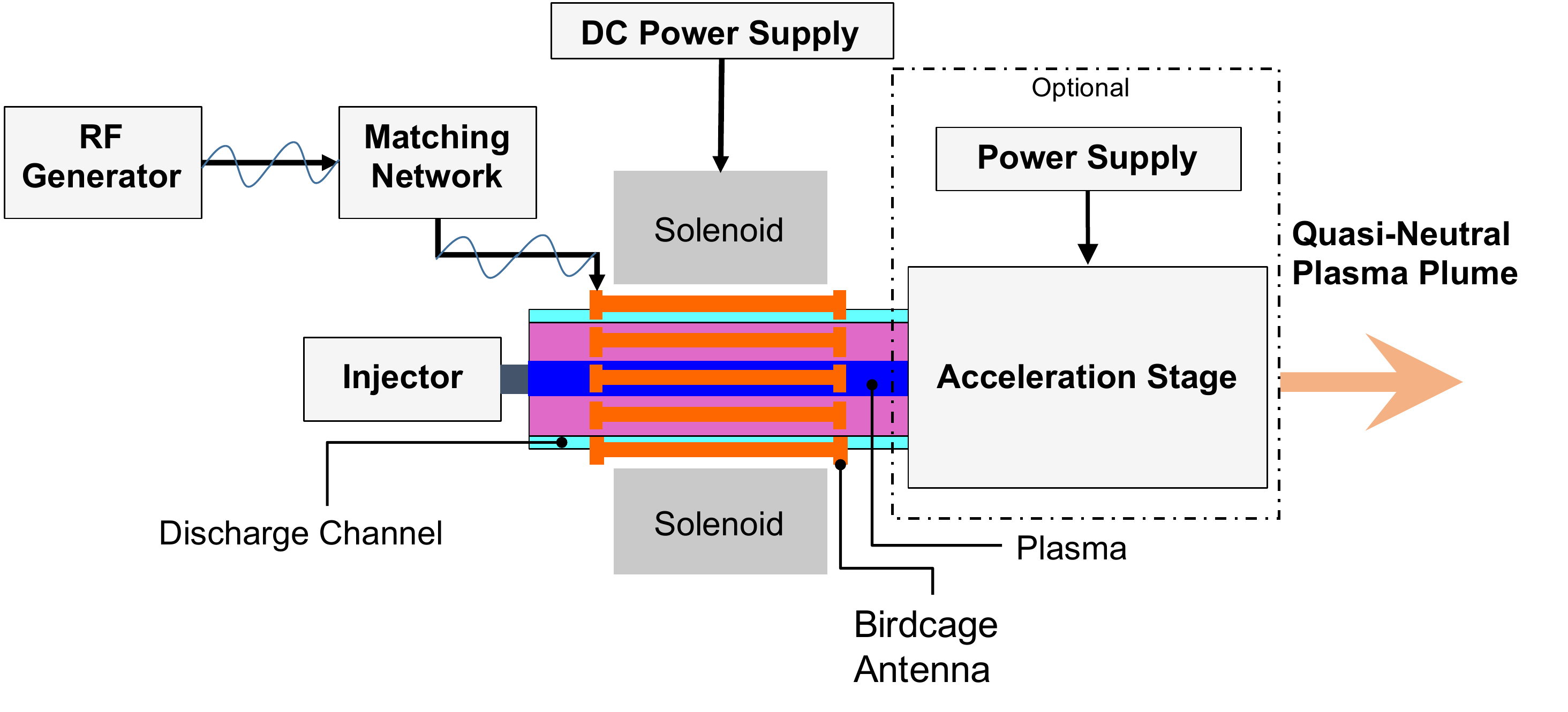}
    \caption{RF Helicon-based Plasma Thruster (IPT) Concept~\cite{romanoacta2}.}
    \label{fig:IPT}
\end{figure}

The thruster concept is based on an antenna fed by a RF power which ionizes and accelerates the propellant. In particular, the antenna surrounds a dielectric discharge channel, in which the propellant is injected. The RF-fed antenna launches electromagnetic waves within the discharge channel exciting the propellant particles: it ionizes them and, in combination with an externally applied magnetic field, that is provided by either a solenoid or a system of permanent magnets, accelerates them through the exhaust producing thrust, see Fig.~\ref{fig:IPT}. 

Furthermore, the applied magnetic field also provides one of the required boundary conditions for helicon waves to develop in the plasma, therefore improving the discharge efficiency~\cite{Chen_2015}.

\subsection{RF Helicon-based Plasma Thruster (IPT)}
The RF Helicon-based plasma thruster (IPT) is a contact-less device based on a novel antenna type, the birdcage antenna. Such an antenna has the advantage of operating at a specific resonance mode that provides not only an already matched load to the RF circuit, but also a linearly polarized magnetic (and electric) field within the IPT's discharge channel. The first property leads to an optimum electrical performance by maximizing the power transfer efficiency, therefore, minimizing losses and heating. The second, the linearly polarized electric and magnetic field, theoretically provides a drift velocity to both ions and electrons along the axial direction due to the $\vec{E}\times\vec{B}$ product that is independent of the particle charge~\cite{romanoacta2,romano_2022}. The IPT is protected/and protects the outer environment from RF waves by means of a Faraday cage made by of brass placed around the birdcage antenna itself. Around the IPT, in the antenna region, a solenoid provides up to $B<\SI{70}{\milli\tesla}$ of static magnetic field at the axis, that enhances the plasma discharge by providing one of the boundary conditions for the formation of helicon waves within the plasma itself~\cite{Chen_2015,romano_2022}. Furthermore, the magnetic field divergence at the exhaust of the IPT provides a magnetic nozzle effect for thrust production, along with the drift velocity given by the birdcage antenna. The RF Helicon-based plasma thruster (IPT) has been simulated, designed, and built for an operational frequency of $f=\SI{40.68}{\mega\hertz}$. 
The experimental set-up is composed by an Advanced Energy CESAR 4040 $f=\SI{40.68}{\mega\hertz}$ RF generator, an Advanced Energy Navigator 4040-L70 automatic matching network, the electromagnet is fed by a GENSYS \SI{750}{\watt} DC power supply, and the IPT itself is flanged to the vacuum chamber. The vacuum chamber has a diameter of $D=\SI{1}{\meter}$ and a length of $L=\SI{2.75}{\meter}$. The vacuum pumping system provides a final base pressure of $p<\SI{0.01}{\pascal}$. 

\begin{figure}[H]
    \centering
    \includegraphics[width=1.1\textwidth]{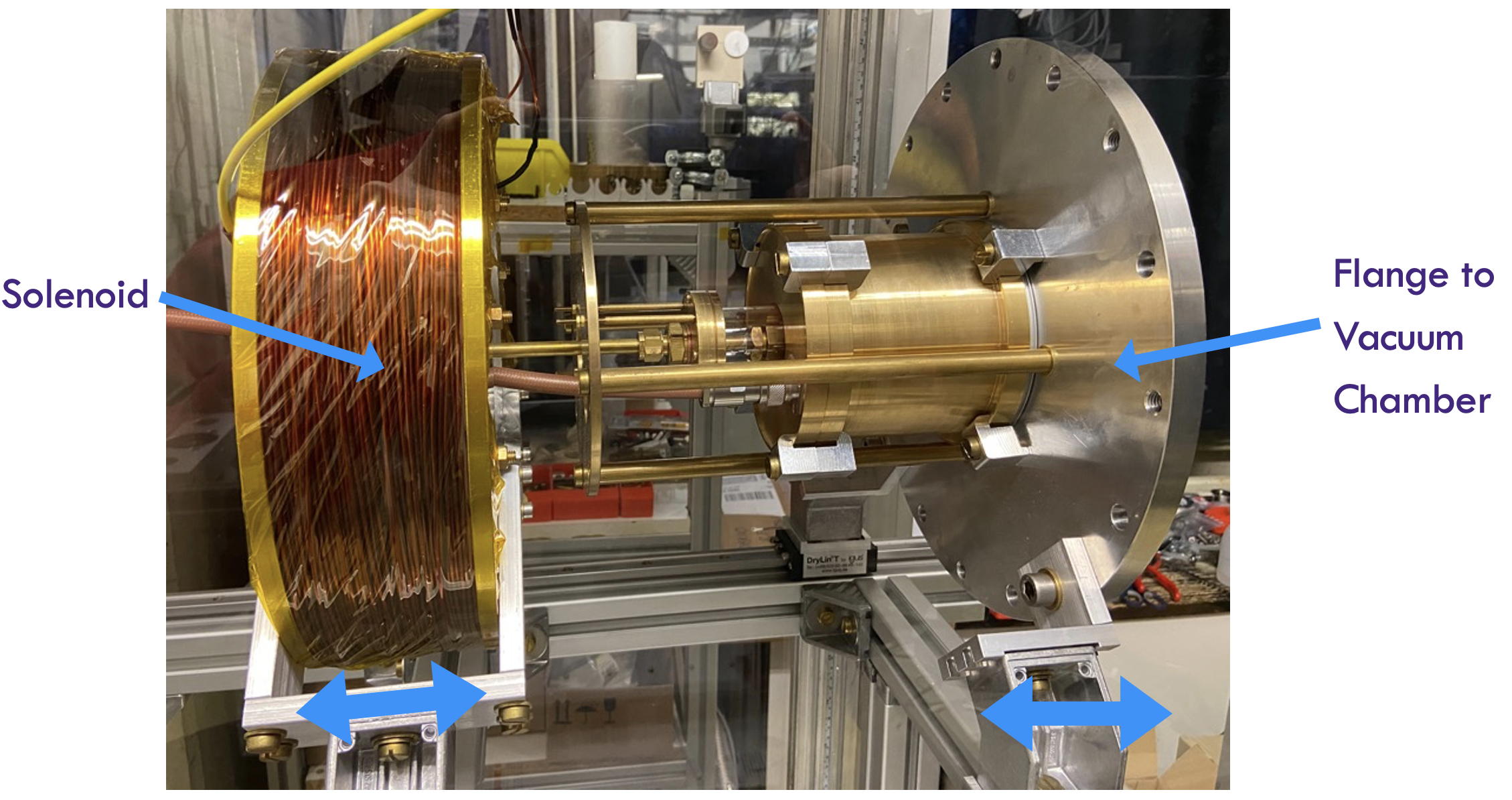}
    \caption{RF Helicon-based Plasma Thruster (IPT) Prototype.}
    \label{fig:IPTlab}
\end{figure}

\begin{figure}[H]
    \centering
    \includegraphics[width=1\textwidth]{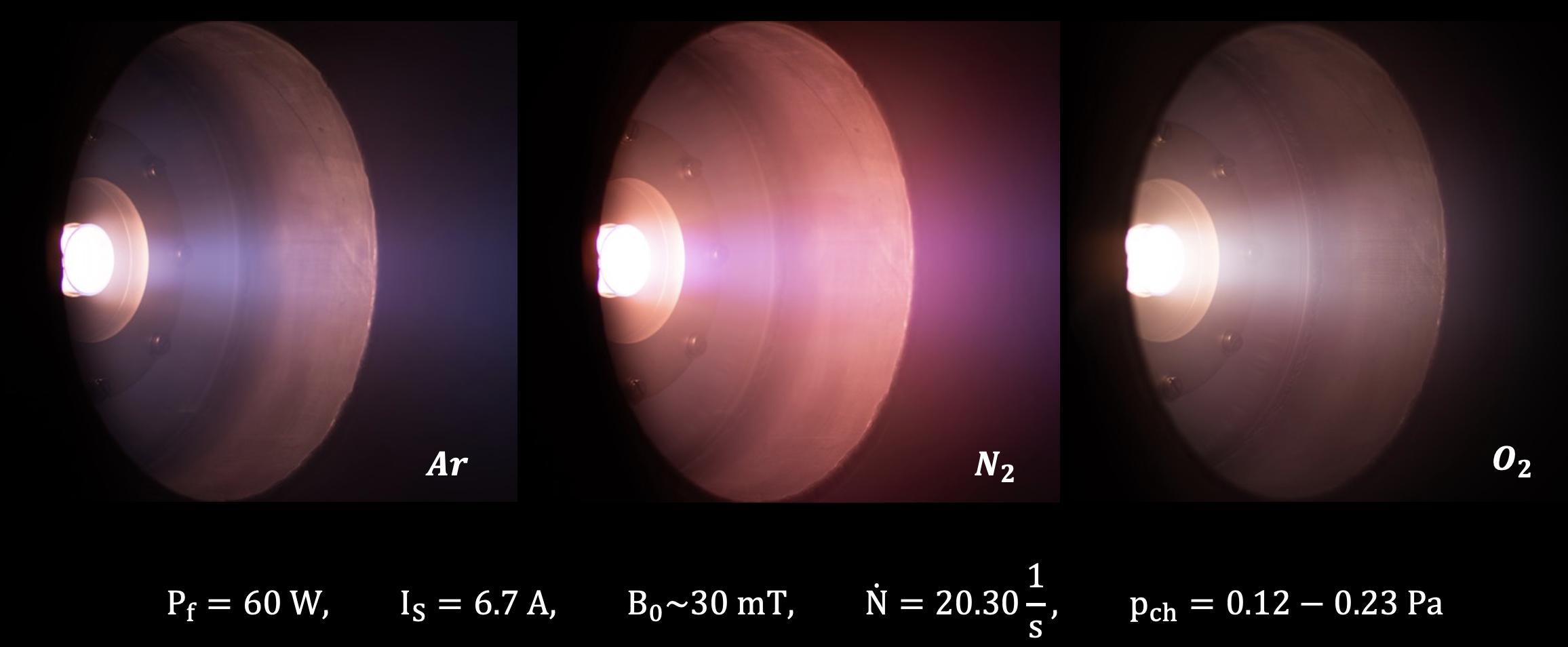}
    \caption{RF Helicon-based Plasma (IPT) in Operation.}
    \label{fig:IPTop}
\end{figure}

The electrical performances have been verified experimentally against simulations in terms of impedance and resonance frequency~\cite{romanoacta2,romano_2022}, and the IPT was successfully demonstrated to operate on \ce{Ar} as well as atmospheric gases such as \ce{N2} and \ce{O2}, in particular ignition, operation at minimum reflected power, and formation of a visible collimated plasma plume in the vacuum chamber after applying the external magnetic field~\cite{romano_2022}. The assembled IPT is shown in Fig.~\ref{fig:IPTlab}, and its operation on \ce{Ar}, \ce{N2}, and \ce{O2} is shown in Fig.~\ref{fig:IPTop}, where the particle flux, power input, and applied magnetic field are equal for the three cases. The three pictures are taken through an optical window with identical camera settings.  

\section{Diagnostic for Thruster Development}
To further develop the IPT, specific diagnostics are required to evaluate the discharge behaviour as well as to determine the thrust produced. First, a scientific evidence of formation of helicon waves is of high interest to highlight the effective operation of the IPT in the helicon regime~\cite{EPFL4,EPFL5}. This can be achieved by employing a 3-axis magnetic-inductive B-dot probe that is currently undergoing calibration at IRS. Such probe allows the measurement of the rotating magnetic field that is typical of helicon waves in the plasma plume~\cite{romanosp2021,romano_2022}. 

\begin{figure}[H]
	\centering
	\includegraphics[width=\textwidth]{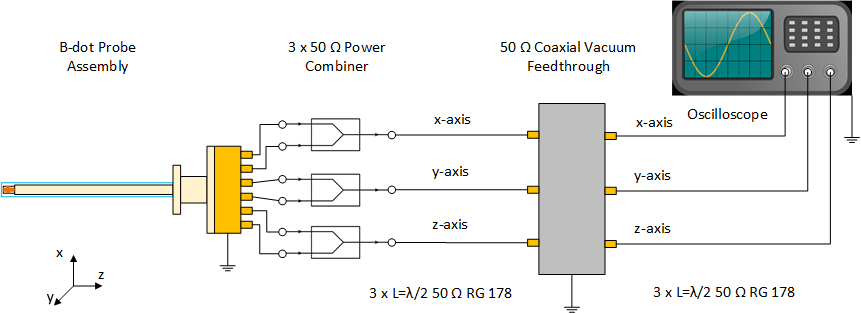}
	\caption{Magnetic Inductive B-dot Probe.}
	\label{fig:Bdot}
\end{figure}

The probe shown in Fig.~\ref{fig:Bdot}, is made of three small $N=5$ turns coils with $\phi=\SI{0.2}{\milli\meter}$ wire, and a coil area respectively of $A_x =\SI{16}{\milli\meter^2}$, $A_y=\SI{25}{\milli\meter^2}$, and $A_z =\SI{36}{\milli\meter^2}$, that can detect the time-varying magnetic fields as they induce a voltage $V$ at an angular frequency $\omega = 2\pi f$, see Eq.~\ref{eq:bdot}. 
\begin{equation}
V=-NA\abs{ \frac{dB_{tot}}{dt} } = -NA \omega \abs{B}
\label{eq:bdot}
\end{equation}
 The probe material is mainly PEEK due to mechanical strength and relatively high melting temperature. An additional quartz cover is added to shield the probe from the direct plasma bombardment. The signal measured by each coil is sent to RF power combiners (one per axis) to remove capacitive pick-up that arises between plasma and coils, see Fig.~\ref{fig:Bdot}. 

To directly measure the produced thrust, isntead, a thrust/force balance is required. 

\begin{figure}[H]
	\centering
	\includegraphics[width=.6\textwidth]{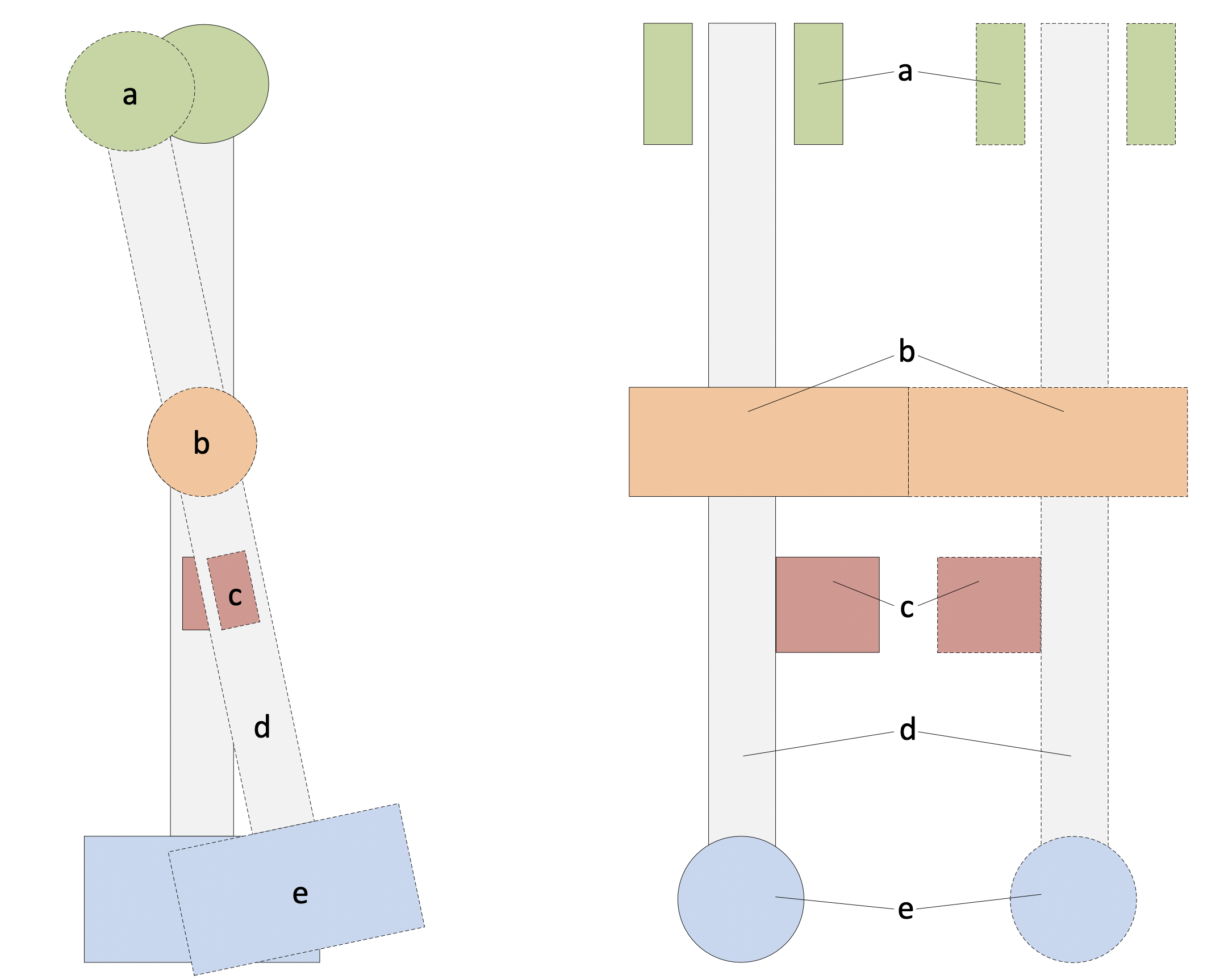}
	\caption{Example of a Pendulum-Type Thrust Balance: a) damper, b) bearing, c) translation sensor, d) pendulum arm structure, e) thruster~\cite{hey2014development}.}
	\label{fig:balance}
\end{figure}

Such device is a mechanical structure that is capable to isolate the thrust force from the external forces, and to translate it into a displacement that can be measured. A schematics of a typical pendulum-type thrust balance set-up is shown within Fig.~\ref{fig:balance}. The pendulum-type is made by the thruster (e) mounted at the bottom of an arm (d) that can oscillate around its center thanks to a bearing (b). On the other side of the arm there is a damper (a), whereas a translation sensor (c) is mounted close to the center of the arm to measure the movement due to the thruster operation. 

To perform such measurement, the current version of the IPT needs to be upgraded and optimized to allow vacuum operation and mounting on the thrust balance, therefore reducing the weight, in particular of the solenoid for the applied magnetic field.

\section{Conclusion}
This article provides the design of the intake and the thruster for an ABEP system. The working principle ABEP intakes based on diffuse and specular-based GSI properties is presented, and highlights the advantages of the specular intake in terms of both intake efficiency, $\eta_c =0.94$, and robustness of its performance to misalignment with the free stream~\cite{romanoacta3}. However, the diffuse intake design providing $\eta_c=0.458$, is still applicable due also to its larger higher-pressure region provided. Finally, the selection always depends on the mission requirements~\cite{romanoacta3}. 

The RF Helicon-based plasma thruster (IPT) is based on RF contact-less technology using an innovative birdcage antenna that allows to minimize power requirement while enhancing thruster operation. The IPT does not have any component in direct contact with the plasma that could erode, critical for ABEP application due to the presence of atomic oxygen (AO), and it does not require a neutralizer as the plasma plume ejected is quasi-neutral. The IPT electrical properties have been experimentally verified~\cite{romanoacta2,romano_2022}, and its proof of concept achieved by operating it on \ce{Ar} as well as on \ce{N2} and \ce{O2}~\cite{romanoiac4,romanosp2021,romano_2022}. 

Currently, a set of sub-scale intake prototypes are being prepared to test in the Rarefied Orbital Aerodynamics Research facility (ROAR) at The University of Manchester~\cite{ROAR} developed within the H2020 DISCOVERER project, creating relevant VLEO AO flow conditions. Also, the future incoming results on the measurements of aerodynamic coefficients from the Satellite for Orbital Aerodynamics Research (SOAR)~\cite{SOAR} will be used as well. The application of real material GSI properties to the simulations is necessary to reach the first full intake prototypes. 

The IPT will undergo a first set of diagnostics to provide a first thrust measurement by a momentum flux probe that is in the development phase. In parallel the detection of helicon waves within the plasma plume is to be performed by applying a magnetic inductive 3-axes B-dot probe that is currently undergoing calibrations~\cite{romanoiac4,romanosp2021,romano_2022}. A full RF-compensated plasma diagnostic measurement is to be performed for fully characterise the IPT. In the near future, a vacuum-capable IPT version is to be developed for later performing direct thrust measurement by means of a thrust pendulum. 

Finally, the intake design work is strongly linked (and iterated) with the thruster design work, to ensure their operation in a common range of pressures, mass flows, and efficiencies. Therefore, both intake and thruster development will mutually feed into each other for the final ABEP complete system design that, once again, is strongly linked to the detailed mission profile as well, but must always allow the operation in a variable environment.

\section*{Acknowledgements}
This project has received funding from the European Union's Horizon 2020 research and innovation program under grant agreement No.~737183. This reflects only the author's view and the European Commission is not responsible for any use that may be made of the information it contains.

\bibliographystyle{elsarticle-num}
\bibliography{bibliography}

\begin{thebibliography}{10}
\expandafter\ifx\csname url\endcsname\relax
  \def\url#1{\texttt{#1}}\fi
\expandafter\ifx\csname urlprefix\endcsname\relax\def\urlprefix{URL }\fi
\expandafter\ifx\csname href\endcsname\relax
  \def\href#1#2{#2} \def\path#1{#1}\fi

\bibitem{vleobenefit}
P.~Roberts, N.~Crisp, V.~{Abrao Oiko}, S.~Edmondson, S.~Haigh, C.~Huyton,
  S.~Livadiotti, R.~Lyons, K.~Smith, L.~Sinpetru, A.~Straker, S.~Worrall,
  F.~Romano, G.~Herdrich, A.~Boxberger, Y.-A. Chan, C.~Traub, S.~Fasoulas,
  K.~Smith, R.~Outlaw, J.~Becedas, R.~Domínguez, D.~González, V.~Hanessian,
  A.~Mølgaard, J.~Nielsen, M.~Bisgaard, D.~Garcia-Almiñana,
  S.~Rodriguez-Donaire, M.~Sureda, D.~Kataria, R.~Villain, J.~S. Perez,
  A.~Conte, B.~Belkouchi, A.~Schwalber, B.~Heißerer, {DISCOVERER} – making
  commercial satellite operations in very low {E}arth orbit a reality, 70th
  International Astronautical Congress, Washington D.C.,
  USA~(IAC-19.C2.6.1x50774) (October 2019).

\bibitem{romanoacta}
F.~Romano, B.~Massuti-Ballester, T.~Binder, G.~Herdrich, S.~Fasoulas,
  T.~Schönherr, System analysis and test-bed for an atmosphere-breathing
  electric propulsion system using an inductive plasma thruster, Acta
  Astronautica 147 (2018) 114 -- 126.
\newblock \href
  {https://doi.org/https://doi.org/10.1016/j.actaastro.2018.03.031}
  {\path{doi:https://doi.org/10.1016/j.actaastro.2018.03.031}}.

\bibitem{di2007ram}
D.~Di~Cara, J.~Gonzalez~del Amo, A.~Santovincenzo, B.~C. Domínguez,
  M.~Arcioni, A.~Caldwell, I.~Roma, {RAM} electric propulsion for low {E}arth
  orbit operation: an {ESA} study, 30th International Electric Propulsion
  Conference 2007, Florence, Italy~(IEPC 2007 162) (September 2007).

\bibitem{presitael1}
G.~Cifali, T.~Misuri, P.~Rossetti, M.~Andrenucci, D.~Valentian, D.~Feili,
  B.~Lotz, Experimental characterization of {HET} and {RIT} with atmospheric
  propellants, 32nd International Electric Propulsion Conference, Wiesbaden,
  Germany~(IEPC 2011 224) (September 2011).

\bibitem{presitael2}
G.~Cifali, D.~Dignani, T.~Misuri, P.~Rossetti, M.~Andrenucci, D.~Valentian,
  F.~Marchandise, D.~Feili, B.~Lotz, Completion of {HET} and {RIT}
  characterization with atmospheric propellants, Space Propulsion 2012,
  Bordeaux, France~(SP2012 2355386) (May 2012).

\bibitem{TSAGI1}
A.~Fitatyev, A.~Golikov, L.~Nosachev, V.~Skvortsov, D.~Padalitsa, Spacecraft
  with air-breathing electric propulsion as the future ultra-speed aircraft,
  71st International Astronautical Congress, Online, The CyberSpace
  Edition~(IAC-20,C4,6,8,x59643) (October 2020).

\bibitem{TSAGI2}
A.~Erofeev, A.~Nikiforov, G.~Popov, M.~Suvorov, S.~Syrin, S.~Khartov,
  Air-breathing ramjet electric propulsion for controlling low-orbit spacecraft
  motion to compensate for aerodynamic drag, Solar System Research 51~(7)
  (2017) 639--645.

\bibitem{filatyev2019control}
A.~S. Filatyev, O.~V. Yanova, The control optimization of low-orbit spacecraft
  with electric ramjet, Acta Astronautica 158 (2019) 23--31.

\bibitem{TSaGI2018a}
M.~Y. Marov, A.~S. Filatyev, Integrated studies of electric propulsion engines
  during flights in the earth's ionosphere, Cosmic Research 56~(2) (2018)
  123--129.
\newblock \href {https://doi.org/10.1134/S0010952518020041}
  {\path{doi:10.1134/S0010952518020041}}.

\bibitem{TSAGI2018ab}
A.~I. Erofeev, Air intake in the transient regime of rarefied gas flow, TsAGI
  Science Journal 49~(2) (2018).

\bibitem{erofeev2017air}
A.~Erofeev, A.~Nikiforov, G.~Popov, M.~Suvorov, S.~Syrin, S.~Khartov,
  Air-breathing ramjet electric propulsion for controlling low-orbit spacecraft
  motion to compensate for aerodynamic drag, Solar System Research 51~(7)
  (2017) 639--645.

\bibitem{kanev2015electro}
S.~Kanev, V.~Petukhov, G.~Popov, S.~Khartov, Electro-rocket ramjet thruster for
  compensating the aerodynamic drag of a low-orbit spacecraft, Russian
  Aeronautics (Iz VUZ) 58~(3) (2015) 286--291.

\bibitem{JAXA}
K.~Fujita, Air intake performance of air breathing ion engines, Journal of the
  Japan Society for Aeronautical and Space Sciences 52~(610) (2004) 514--521.
\newblock \href {https://doi.org/10.2322/jjsass.52.514}
  {\path{doi:10.2322/jjsass.52.514}}.

\bibitem{JAXA2}
K.~Fujita, Air-intake performance estimation of air-breathing ion engines,
  Transactions of the Japan Society of Mechanical Engineers. B 70~(700) (2004)
  3038--3044.

\bibitem{JAXA3}
Y.~Hisamoto, K.~Nishiyama, H.~Kuninaka, Development statue of atomic oxygen
  simulator for air breathing ion engine, in: 32nd Intern. Electric Propulsion
  Conference. IEPC 2011 (Wiesbaden, Germany, Sept. 11-15 2011): Proc, 2011.

\bibitem{JAXA4}
Y.~Hisamoto, K.~Nishiyama, H.~Kuninaka, Design of air intake for air breathing
  ion engine, in: 63rd International Astronautical Congress, no.
  IAC-12,C4,4,10,x14578, 2012.

\bibitem{JAXA5}
M.~Tagawa, K.~Yokota, K.~Nishiyama, H.~Kuninaka, Y.~Yoshizawa, D.~Yamamoto,
  T.~Tsuboi, Experimental study of air breathing ion engine using laser
  detonation beam source, Journal of Propulsion and Power 29~(3) (2013)
  501--506.
\newblock \href {https://doi.org/10.2514/1.B34530}
  {\path{doi:10.2514/1.B34530}}.

\bibitem{busek}
K.~Hohman, Atmospheric breathing electric thruster for planetary exploration,
  {B}usek {C}o. {I}nc. 11 (2012) 01760--1023.

\bibitem{busek2}
K.~Hohman, Atmospheric breathing electric thruster for planetary exploration,
  Presented as the NIAC Spring Symposium, Presentation, 2012.

\bibitem{SITAEL2015}
S.~Barral, G.~Cifali, R.~Albertoni, M.~Andrenucci, L.~Walpot, Conceptual design
  of an air-breathing electric propulsion system, in: Joint Conf. of 30th Int.
  Symp. on Space Technology and Science, 34th Int. Electric Propulsion Conf.
  and 6th Nano-Satellite Symp, 2015.

\bibitem{SITAEL2016}
G.~Cifali, T.~Andrenussi, V.~Giannetti, A.~Leoprini, A.~Rossodivita,
  M.~Andrenucci, S.~Barral, J.~Longo, L.~Walpot, Experimental validation of a
  {RAM-EP} concept based on {H}all-effect thruster, Space Propulsion 2016,
  Rome, Italy~(SP2016 3125202) (May 2016).

\bibitem{SITAEL2017}
T.~Andreussi, G.~Cifali, V.~Giannetti, A.~Piragino, E.~Ferrato, A.~Rossodivita,
  M.~Andrenucci, J.~Longo, L.~Walpot, Development and experimental validation
  of a {H}all-effect thruster {RAM-EP} concept, 35th International Electric
  Propulsion Conference, Atlanta, USA~(IEPC 2017 377) (October 2017).

\bibitem{SITAEL2019a}
T.~Andreussi, E.~Ferrato, V.~Giannetti, A.~Piragino, C.~A. Paissoni, G.~Cifali,
  M.~Andrenucci, Development status and way forward of {SITAEL}'s air-breathing
  electric propulsion engine, in: AIAA Propulsion and Energy 2019 Forum, 3996,
  2019.

\bibitem{SITAEL2019b}
E.~Ferrato, V.~Giannetti, A.~Piragino, M.~Andrenucci, T.~Andreussi, C.~A.
  Paissoni, Development roadmap of {SITAEL}'s {RAM-EP} system, 36th
  International Electric Propulsion Conference, Vienna, Austria~(IEPC 2019 886)
  (September 2019).

\bibitem{bauman}
D.~V. Dukhopelnikov, V.~A. Riazanov, S.~O. Shilov, D.~S. Manegin, R.~A.
  Sokolov, Investigation of the laboratory model of a thruster with anode layer
  operating with air and nitrogen-oxygen mixture, AIP Conference Proceedings
  2318~(1) (2021) 040006.
\newblock \href {https://doi.org/10.1063/5.0036251}
  {\path{doi:10.1063/5.0036251}}.

\bibitem{shabshelowitz2013study}
A.~Shabshelowitz, Study of {RF} plasma technology applied to air-breathing
  electric propulsion, Ph.D. thesis, University of Michigan (2013).

\bibitem{romanoacta2}
F.~Romano, Y.-A. Chan, G.~Herdrich, C.~Traub, S.~Fasoulas, P.~Roberts,
  K.~Smith, S.~Edmondson, S.~Haigh, N.~Crisp, V.~Oiko, S.~Worrall,
  S.~Livadiotti, C.~Huyton, L.~Sinpetru, A.~Straker, J.~Becedas, R.~Domínguez,
  D.~González, V.~Cañas, V.~Sulliotti-Linner, V.~Hanessian, A.~Mølgaard,
  J.~Nielsen, M.~Bisgaard, D.~Garcia-Almiñana, S.~Rodriguez-Donaire,
  M.~Sureda, D.~Kataria, R.~Outlaw, R.~Villain, J.~Perez, A.~Conte,
  B.~Belkouchi, A.~Schwalber, B.~Heißerer, {RF} {H}elicon-based {I}nductive
  {P}lasma {T}hruster ({IPT}) {D}esign for an {A}tmosphere-{B}reathing
  {E}lectric {P}ropulsion system ({ABEP}), Acta Astronautica 176 (2020) 476 --
  483.
\newblock \href
  {https://doi.org/https://doi.org/10.1016/j.actaastro.2020.07.008}
  {\path{doi:https://doi.org/10.1016/j.actaastro.2020.07.008}}.

\bibitem{VLEO_LEOMANNI2017}
M.~Leomanni, A.~Garulli, A.~Giannitrapani, F.~Scortecci, Propulsion options for
  very low earth orbit microsatellites, Acta Astronautica 133 (2017) 444--454.
\newblock \href
  {https://doi.org/https://doi.org/10.1016/j.actaastro.2016.11.001}
  {\path{doi:https://doi.org/10.1016/j.actaastro.2016.11.001}}.

\bibitem{6945885}
T.~Sch\"{o}nherr, K.~Komurasaki, F.~Romano, B.~Massuti-Ballester, G.~Herdrich,
  Analysis of atmosphere-breathing electric propulsion, Plasma Science, IEEE
  Transactions on 43~(1) (2015) 287--294.
\newblock \href {https://doi.org/10.1109/TPS.2014.2364053}
  {\path{doi:10.1109/TPS.2014.2364053}}.

\bibitem{romano_2022}
F.~Romano, RF {H}elicon {P}lasma {T}hruster for an {A}tmosphere-{B}reathing
  {E}lectric {P}ropulsion {S}ystem ({ABEP}), Verlag Dr. Hut, 2022.

\bibitem{vaidya2022development}
S.~Vaidya, C.~Traub, F.~Romano, G.~Herdrich, Y.-A. Chan, S.~Fasoulas,
  P.~Roberts, N.~Crisp, S.~Edmondson, S.~Haigh, et~al., Development and
  analysis of novel mission scenarios based on atmosphere-breathing electric
  propulsion ({ABEP}), CEAS Space Journal (2022) 1--18\href
  {https://doi.org/10.1007/s12567-022-00436-1}
  {\path{doi:10.1007/s12567-022-00436-1}}.

\bibitem{romanoacta3}
F.~Romano, J.~Espinosa-Orozco, M.~Pfeiffer, G.~Herdrich, N.~Crisp, P.~Roberts,
  B.~Holmes, S.~Edmondson, S.~Haigh, S.~Livadiotti, A.~Macario-Rojas, V.~Oiko,
  L.~Sinpetru, K.~Smith, J.~Becedas, V.~Sulliotti-Linner, M.~Bisgaard,
  S.~Christensen, V.~Hanessian, T.~K. Jensen, J.~Nielsen, Y.-A. Chan,
  S.~Fasoulas, C.~Traub, D.~García-Almiñana, S.~Rodríguez-Donaire,
  M.~Sureda, D.~Kataria, B.~Belkouchi, A.~Conte, S.~Seminari, R.~Villain,
  Intake design for an {A}tmosphere-{B}reathing {E}lectric {P}ropulsion system
  ({ABEP}), Acta Astronautica 187 (2021) 225--235.
\newblock \href {https://doi.org/10.1016/j.actaastro.2021.06.033}
  {\path{doi:10.1016/j.actaastro.2021.06.033}}.

\bibitem{LIVADIOTTI2020}
S.~Livadiotti, N.~H. Crisp, P.~C. Roberts, S.~D. Worrall, V.~T. Oiko,
  S.~Edmondson, S.~J. Haigh, C.~Huyton, K.~L. Smith, L.~A. Sinpetru, B.~E.
  Holmes, J.~Becedas, R.~M. Domínguez, V.~Cañas, S.~Christensen,
  A.~Mølgaard, J.~Nielsen, M.~Bisgaard, Y.-A. Chan, G.~H. Herdrich, F.~Romano,
  S.~Fasoulas, C.~Traub, D.~Garcia-Almiñana, S.~Rodriguez-Donaire, M.~Sureda,
  D.~Kataria, B.~Belkouchi, A.~Conte, J.~S. Perez, R.~Villain, R.~Outlaw, A
  review of gas-surface interaction models for orbital aerodynamics
  applications, Progress in Aerospace Sciences 119 (2020) 100675.
\newblock \href {https://doi.org/10.1016/j.paerosci.2020.100675}
  {\path{doi:10.1016/j.paerosci.2020.100675}}.

\bibitem{Munz2014662}
C.-D. Munz, M.~Auweter-Kurtz, S.~Fasoulas, A.~Mirza, P.~Ortwein, M.~Pfeiffer,
  T.~Stindl, Coupled particle-in-cell and direct simulation monte carlo method
  for simulating reactive plasma flows, Comptes Rendus Mécanique 342~(10–11)
  (2014) 662 -- 670, theoretical and numerical approaches for Vlasov-maxwell
  equations.
\newblock \href {https://doi.org/http://dx.doi.org/10.1016/j.crme.2014.07.005}
  {\path{doi:http://dx.doi.org/10.1016/j.crme.2014.07.005}}.

\bibitem{PICLAS}
S.~Fasoulas, C.-D. Munz, M.~Pfeiffer, J.~Beyer, T.~Binder, S.~Copplestone,
  A.~Mirza, P.~Nizenkov, P.~Ortwein, W.~Reschke, Combining particle-in-cell and
  direct simulation monte carlo for the simulation of reactive plasma flows,
  Physics of Fluids 31~(7) (2019) 072006.
\newblock \href {https://doi.org/10.1063/1.5097638}
  {\path{doi:10.1063/1.5097638}}.

\bibitem{Chen_2015}
F.~F. Chen, Helicon discharges and sources: a review, Plasma Sources Science
  and Technology 24~(1) (2015) 014001.
\newblock \href {https://doi.org/10.1088/0963-0252/24/1/014001}
  {\path{doi:10.1088/0963-0252/24/1/014001}}.

\bibitem{EPFL4}
R.~Jacquier, R.~Agnello, B.~P. Duteil, P.~Guittienne, A.~Howling,
  G.~Plyushchev, C.~Marini, A.~Simonin, I.~Morgal, S.~Bechu, et~al., First
  {B}-dot measurements in the {RAID} device, an alternative negative ion source
  for demo neutral beams, Fusion Engineering and Design (2019).
\newblock \href
  {https://doi.org/https://doi.org/10.1016/j.fusengdes.2019.02.025}
  {\path{doi:https://doi.org/10.1016/j.fusengdes.2019.02.025}}.

\bibitem{EPFL5}
P.~Guittienne, R.~Jacquier, B.~P. Duteil, A.~A. Howling, R.~Agnello, I.~Furno,
  Helicon wave plasma generated by a resonant birdcage antenna: magnetic field
  measurements and analysis in the {RAID} linear device, Plasma Sources Science
  and Technology 30~(7) (2021) 075023.
\newblock \href {https://doi.org/10.1088/1361-6595/ac0da3}
  {\path{doi:10.1088/1361-6595/ac0da3}}.

\bibitem{romanosp2021}
F.~Romano, Y.-A. Chan, G.~Herdrich, C.~Traub, S.~Fasoulas, P.~Roberts,
  N.~Crisp, B.~Holmes, S.~Edmonson, S.~Haigh, S.~Livadiotti, A.~Macario-Rojes,
  V.~Oiko, L.~Sinpetru, K.~Smith, J.~Becedas, V.~Sulliotti-Linner, M.~Bisgaard,
  S.~Christensen, V.~Hanessian, T.~K. Jensen, J.~Nielsen, D.~Garcia-Almiñana,
  M.~G.-B.~S. Rodrigue-Donaire, M.~Sureda, D.~Kataria, B.~Belkouchi, A.~Conte,
  S.~Seminari., Design, set-up, and first ignition of the {RF} {H}elicon-based
  {P}lasma {T}hruster, Space Propulsion 2021, Online~(SP2021 00247) (March
  2021).

\bibitem{hey2014development}
F.~G. Hey, A.~Keller, C.~Braxmaier, M.~Tajmar, U.~Johann, D.~Weise, Development
  of a highly precise micronewton thrust balance, IEEE Transactions on Plasma
  Science 43~(1) (2014) 234--239.

\bibitem{romanoiac4}
F.~Romano, G.~Herdrich, , Y.-A. Chan, P.~C. Roberts, N.~H. Crisp,
  A.~Macario~Rojas, S.~Edmondson, S.~Haigh, B.~Holmes, S.~Livadiotti, V.~A.
  Oiko, K.~Smith, L.~Sinpetru, J.~Becedas, R.~Dom{\'i}nguez,
  V.~Sulliotti-Linne, S.~Christensen, T.~{Kauffman Jensen}, J.~Nielsen,
  M.~Bisgaard, S.~Fasoulas, C.~Traub, D.~Garcia-Almi{\~n}ana,
  M.~Garcia-Berenguer, S.~Rodriguez-Donaire, M.~Sureda, D.~Kataria,
  B.~Belkouchi, A.~Conte, S.~Seminari, R.~Villain, {RF} helicon-based plasma
  thruster {(IPT)}: Design, set-up, and first ignition, 71st International
  Astronautical Congress, Online, The CyberSpace
  Edition~(IAC-20,C4,5,11,x58032) (October 2020).

\bibitem{ROAR}
V.~{Abrao Oiko}, P.~Roberts, A.~{Macario Rojas}, S.~Edmondson, S.~Haigh,
  B.~Holmes, S.~Livadiotti, N.~Crisp, K.~Smith, L.~Sinpetru, J.~Becedas,
  R.~Dom{\'i}nguez, V.~Sulliotti-Linne, S.~Christensen, T.~{Kauffman Jensen},
  J.~Nielsen, M.~Bisgaard, Y.-A. Chan, G.~Herdrich, F.~Romano, S.~Fasoulas,
  C.~Traub, D.~Garcia-Almi{\~n}ana, M.~Garcia-Berenguer, S.~Rodriguez-Donaire,
  M.~Sureda, D.~Kataria, B.~Belkouchi, A.~Conte, S.~Seminari, R.~Villain,
  Ground-based experimental facility for orbital aerodynamics research: Design,
  construction and characterisation, in: 71st International Astronautical
  Congress (IAC) – The CyberSpace Edition, 12-14 October 2020, 2020.

\bibitem{SOAR}
N.~Crisp, P.~Roberts, S.~Livadiotti, A.~{Macario Rojas}, V.~Oiko, S.~Edmondson,
  S.~Haigh, B.~Holmes, L.~Sinpetru, K.~Smith, J.~Becedas, R.~Domínguez,
  V.~Sulliotti-Linner, S.~Christensen, J.~Nielsen, M.~Bisgaard, Y.-A. Chan,
  S.~Fasoulas, G.~Herdrich, F.~Romano, C.~Traub, D.~García-Almiñana,
  S.~Rodríguez-Donaire, M.~Sureda, D.~Kataria, B.~Belkouchi, A.~Conte,
  S.~Seminari, R.~Villain, In-orbit aerodynamic coefficient measurements using
  {SOAR} ({S}atellite for {O}rbital {A}erodynamics {R}esearch), Acta
  Astronautica 180 (2021) 85--99.
\newblock \href
  {https://doi.org/https://doi.org/10.1016/j.actaastro.2020.12.024}
  {\path{doi:https://doi.org/10.1016/j.actaastro.2020.12.024}}.

\end{thebibliography}

\end{document}